\documentclass[12pt]{article}
\usepackage{bm}
\usepackage{graphicx}
\begin{document}

\title{\textbf{Semiclassical Mechanics of Rotons}}
\author{T. M. Sanders, Jr.
\thanks{\emph{Author's address:}
    H. M. Randall Laboratory, University 
  of Michigan, Ann Arbor, MI 48109-1120, USA.}}
\date{\today}
\maketitle

\begin{abstract}
  
  The elementary excitations in superfluid liquid ${}^4\textrm{He}$
  named rotons have an unusual dispersion curve. The energy is an
  approximately quadratic function of $(p-p_0)$, the difference
  between the magnitude of the momentum $p$ and a characteristic value
  $p_0$. As a result, while for $p>p_0$ a roton has its (group)
  velocity parallel to its momentum, when $p<p_0$ the velocity and
  momentum are antiparallel. When $p=p_0$, the roton has non-zero
  momentum but zero velocity. These kinematic properties lead to
  unusual trajectories when rotons scatter or experience external
  forces. This paper examines this behavior in the classical (ray
  optics) limit, where the roton wavelength is small compared with all
  other dimensions. Several experiments illustrate these effects. The
  examples are interesting in themselves, and also offer
  unconventional pedagogical possibilities.

\end{abstract}

\section{Introduction}

\subsection{Background}

The concept of elementary excitations in superfluid liquid helium
(${}^4\textrm{He}$) was introduced by L.~D.~Landau in
1941.~\cite{Landau:1941} The essential idea of Landau is that the
low-lying excited states of the macroscopic system have energies $E$
($E_0$ is the ground state energy.)  and momentum $\vec{P}$ which can
be expressed in the form

\[
 E = E_0 + \sum_{\vec{p}} n(\vec{p}) \varepsilon(\vec{p})
\]
\[
 \vec{P} = \sum_{\vec{p}} n(\vec{p}) \vec{p}. 
\]

That is, the system of strongly-interacting particles (He atoms) is
replaced by a collection of non (or weakly) interacting ``elementary
excitations'' or ``quasiparticles.''  In Landau's paper of 1941, he
suggested a form for the dispersion curve (the relation between energy
$\varepsilon $ and momentum $\vec{p}$) for these entities. For an
isotropic system, such as a liquid, the energy can depend only on $p,$
the magnitude of the momentum. The 1941 dispersion curve has two
branches: There is an ``acoustic'' branch with $\varepsilon(p) = cp$,
representing longitudinal phonons ($c$ is the speed of sound.), and a
second branch with $\varepsilon(p) = p^2/2\mu + \Delta$. Landau
called these elementary excitations
``rotons.''\footnote{Landau~\cite{Landau:1941} writes that ``This name
was suggested by I.~E.~Tamm''} The situation seems reminiscent of that
in a solid, with its acoustical and optical branches. There is,
however, an important difference: The liquid has complete
translational invariance (not just against finite translations). Thus
the momentum of the elementary excitations is not just a quantum
number like crystal momentum, but is the usual physical
momentum. Landau evaluated the two free parameters $\Delta$ and $\mu$
by fitting measured values of the specific heat of the liquid, and
used the model to calculate the normal and superfluid fractions as
functions of the temperature.

In his brief 1947 paper,~\cite{Landau:1947} Landau\footnote{Landau's
papers of 1941 and 1947 are reprinted in several books, such
as~\cite{Landau:1965} and~\cite{Khalatnikov:1988}}
modified the relationship $\varepsilon (p)$ in order to account for
some experimental data on the propagation of second sound. The curve
he drew in 1947 agrees remarkably well with modern determinations of
$\varepsilon (p),$ using inelastic neutron scattering. The two
branches are replaced by a single branch which has a linear part, a
maximum, and a minimum: We call the elementary excitations near the
origin ``phonons'', those near the minimum ``rotons.'' These two
parts of the spectrum are dominant in determining the
low-temperature properties of liquid
helium. Donnelly~\cite{Donnelly:1997a} has written a recent
introduction to the subject with interesting historical perspective
and some discussion of the microscopic nature of the roton, a subject
beyond the scope of the present article.

Landau emphasized the importance of the finite slope of the dispersion
curve at the origin---in distinction to the behavior of the Ideal Bose
Gas. A consequence of this finite slope is a ``critical velocity'',
below which the flow of the helium is dissipationless. In the simplest
case the primary dissipation process would be roton emission. With
rare exceptions~\cite{Ellis:1985} other dissipation mechanisms occur at
velocities below that required for roton emission.

The reasons why there are no other states near the ground state of an
interacting Bose system are considered in the papers of
Feynman~\cite{Feynman:1955} in the 1950s.

\subsection{Current interest}

Since the development of the dilution refrigerator, it has become
practical to perform experiments on ``ballistic rotons.'' That is, at
sufficiently low temperature the mean free path of a roton becomes as
long as the size of the apparatus. Rotons (and phonons) are produced
by a ``heat pulse'' in the liquid or, in some cases, by helium atoms
incident on the surface of the liquid from the (low-density) vapor
above it. We will consider such experiments in
sections~\ref{sec:surface}~and~\ref{sec:expers}. 

\section{The dispersion curve}

\begin{figure}[htbp]
  \begin{center}
    \includegraphics{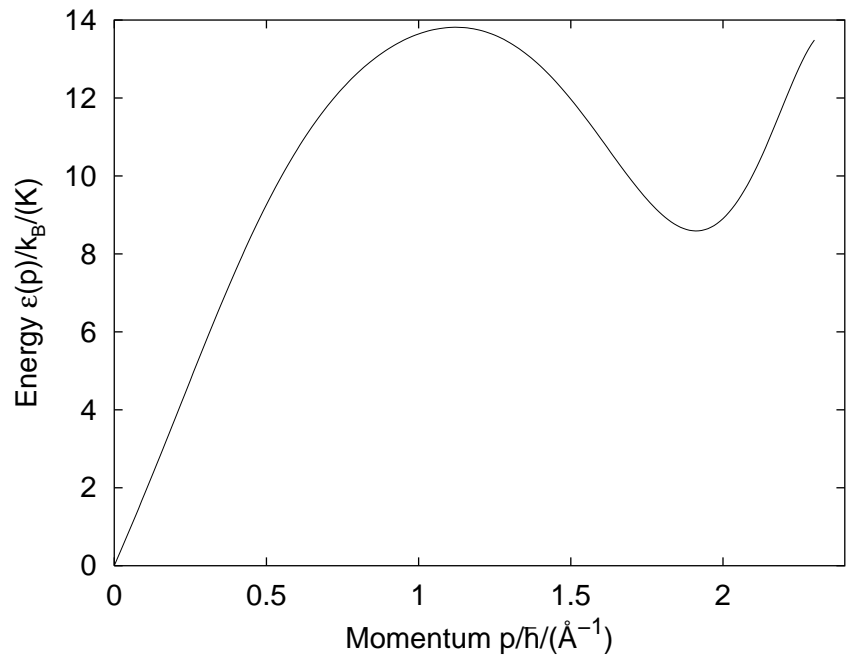}
  \end{center}
  \caption{Dispersion curve for elementary excitations in
    liquid helium, calculated from the polynomial fit in reference~{\protect
      \cite{Brooks:1977}}. The roton region is near the minimum.}
  \label{fig:landaucurve}
\end{figure}

The dispersion curve for elementary excitations in superfluid
${}^4\textrm{He}$, determined by inelastic neutron scattering
\cite{Donnelly:1981}. is shown in figure~\ref{fig:landaucurve}. The region
near the minimum is usually described by the equation

\begin{equation}
\varepsilon(p)=(p-p_0)^2/2\mu + \Delta.\label{irr}
\end{equation}

At low temperature and pressure, approximate values for the parameters are:
$q_0=p_0/\hbar=1.92\,\textrm{\AA}^{-1},\ \Delta/k_B=8.62$ K, $\mu=m_4/6$.
The function $\varepsilon(p)$ can be considered a Hamiltonian for the free
roton, since $\partial\varepsilon/\partial p_i=v_i$ is a component of the
(group) velocity of a roton wave packet. Using the expression~(\ref{irr}),
we have $\vec{v}=\hat{p}(p-p_0)/\mu$.

\subsection{Statistical Thermodynamics}

We recapitulate some results which we will later find useful. Since we
will be interested in temperatures less than or of order 1\,K, we will
always have $\varepsilon(p)\ge\Delta>>k_BT,$ whence
$\exp(-\varepsilon(p)/k_BT)=\exp(-\beta\varepsilon(p))<<1$ will always
be valid. (The symbol $\beta$ stands, as is usual, for $1/k_BT$.) This
permits considerable simplification of various equations.
Furthermore, all integrand functions of $p$ will have contributions
peaked near $p=p_0$. Thus, we transform integrals over $p$ over the
range from zero to infinity to integrals over $(p-p_0)$ from $-\infty$
to $\infty $.

\noindent The expected number of rotons in a given mode $\vec{p}$:

\[
\left < n(\vec{p})\right > = (\exp(\beta\varepsilon(p))-1)^{-1}
\approx \exp(-\beta\varepsilon(p)).
\]

\noindent The number density of rotons:

\[
n_\rho \equiv N_\rho/V=(2\pi\hbar)^{-3}\int_0^\infty
{\exp(-\beta\varepsilon(p))4\pi p^2dp}
\]

\noindent The Helmholtz Free Energy:

\[
F_\rho = -k_BTV/(2\pi\hbar)^3\int_0^\infty{\exp(-\beta\varepsilon(p))4\pi p^2dp}
\]

\noindent The average ``kinetic'' energy per roton:

\begin{eqnarray}
\left < (p-p_0)^2/2\mu \right > &=&\left < \mu v^2/2 \right >
\nonumber \\ 
& =&  - \partial/\partial \beta \left (\log
\int_0^\infty{\exp \left ( -\beta (p-p_0)^2/2\mu \right ) p^2dp}\right )
\nonumber \\
& =& \frac{1}{2}k_BT \nonumber 
\end{eqnarray}

\noindent The pressure due to the roton gas:

\begin{equation}
\label{p_rho}
  p_\rho=-\left ( \partial F/\partial V \right ) _T \cong -F/V = n_\rho
  k_BT = n_\rho \mu \left < v^2 \right >
\end{equation}

\section{Free rotons}

\begin{figure}[ht]
  \begin{center}
    \includegraphics{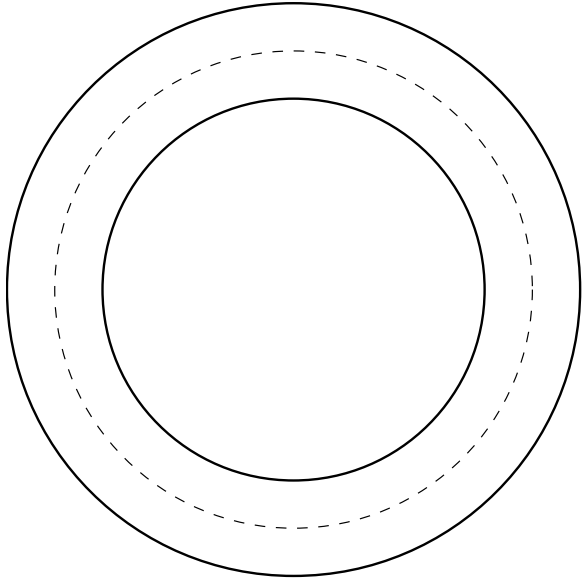}
  \end{center}
\caption{Surfaces of constant energy
  in momentum space for $\varepsilon=\Delta$ (dotted), and a higher
  energy (solid).}\label{fig:isoenergy}
\end{figure}

It is worth repeating that the expression for the velocity has the
curious property that $\vec{v}$ is parallel to $\vec{p}$ when $p>p_0$,
but \emph{antiparallel} when $p$ is below the minimum. We will refer to
rotons with momentum above the minimum as $R^+$ or \emph{ordinary},
those with momentum below as $R^-$ or \emph{extraordinary}. The surface
$\varepsilon(p)=\Delta$ is a sphere of radius $p_0$ in momentum space.
For an energy greater than $\Delta$ (but below the maximum) the
surface has two concentric spherical sheets, one with $p>p_0$, on
which the rotons are ordinary, the other with $p<p_0$, describing
extraordinary rotons. Figure~\ref{fig:isoenergy} shows a
two-dimensional representation. We will often denote the two
``conjugate'' values of momentum with equal energy as $p$ and
$\tilde{p}$.

\section{Rotons in a velocity field}
\label{sec:vel}

A somewhat elusive argument, due to Landau, yields the result that
to produce an elementary excitation of momentum $\vec{p}$ in a fluid
moving at velocity $\vec{v}$ requires energy
\begin{equation}
\label{doppler}
\varepsilon ^{(\vec{v})} (\vec{p}) = \varepsilon (\vec{p})
+ \vec{p}\cdot \vec{v},
\end{equation}
where $\varepsilon (\vec{p})$ is the energy of the excitation in the
stationary fluid.
In a slowly-varying velocity field, we add a term to the hamiltonian
\begin{equation}
\label{pdotv}
\delta \varepsilon (\vec{p}) = \vec{p}\cdot \vec{v}(\vec{r}).
\end{equation}

A quasiparticle of a particular momentum magnitude has higher energy
if its momentum points ``downstream'' than if it is directed
``upstream''. Thus, in thermal equilibrium, the excitations will
reduce the momentum of a moving superfluid. Landau uses this argument to
calculate the normal-fluid density. In section~\ref{sec:vortex}, we
will apply it to the problem of roton motion near a quantized vortex.

We present a rather pedestrian but (we hope) straightforward derivation of
this familiar result in Appendix~\ref{sec:doppler}

\section{Rotons in a force field}

The roton parameters $\Delta,\ p_0,$ and $\mu$ can vary with position,
since they are known to depend on the density of the liquid. The
``energy gap'' $\Delta,$ in particular, decreases from 8.62\,K at the
vapor pressure to approximately 7.2\,K at a pressure of 25 atm
\cite{Donnelly:1981}. If the roton parameters vary sufficiently
slowly compared with the roton wavelength, the roton will follow a
classical trajectory. Since $q_0 = p_0/\hbar \approx 2
\textrm{\AA}^{-1},$ we have $\lambda \approx 3\textrm{\AA} ;$ a
spatial variation of $\Delta $ on a scale large compared with
$3\textrm{\AA} $ will appear like a classical potential energy in
$\varepsilon(p)\equiv {\mathcal H}(\vec{p},\vec{r})$. In the case in which
only variation in $\Delta$ is considered, we have a Hamiltonian
consisting of a ``kinetic'' energy $T=(p - p_0)^2/2\mu,$ and a
``potential'' energy $V(\vec{r})=\Delta (\vec{r})$.

\subsection{A constant force}

An interesting illustration of the consequences of the unusual dispersion
relation is provided by considering the motion of rotons in a uniform
force field, i.e.\ when $\Delta$ is a linear function of a single
coordinate, say $y$. If we write $f=-d\Delta/dy,$ then Hamilton's
equations take the form

\begin{eqnarray}
\dot{p_x} & = & - \partial{\mathcal H}/\partial{x} \nonumber \\
& = & 0 \nonumber  \\
\dot{p_y} & = & - \partial{\mathcal H}/\partial{y} \nonumber \\
& =& f \nonumber  \\
\dot{x} & = & \partial{\mathcal H}/\partial{p_x}  \nonumber \\
& = & (p_x/p) \times dT/dp \nonumber  \\
\dot{y} & = & \partial{\mathcal H}/\partial{p_y}  \nonumber \\
& = & (p_y/p) \times dT/dp. \nonumber 
\end{eqnarray}

\begin{figure}[ht]
  \begin{center}
    \includegraphics{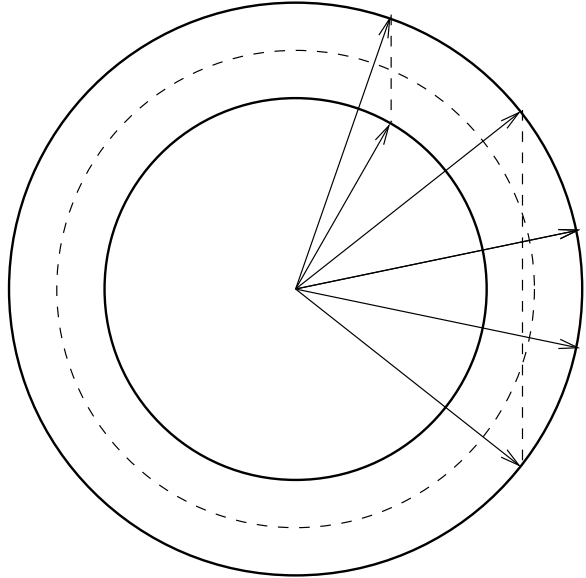}
  \end{center}
\caption{Momentum space diagram
  showing initial and final momenta corresponding to reflection of
  rotons by a force in the $-y$ direction.}\label{fig:pipf}
\end{figure}

The equations for $\dot{\vec{p}}$ can be integrated immediately when
the force $f$ is constant. We obtain $p_x=p_{xi},\ p_y=p_{yi}+ft$. On
a momentum-space diagram such as that of figure~\ref{fig:isoenergy} the
motion is parallel to the $p_y$ axis, in a direction given by the sign
of $f$. For $f<0,$ we show the various possible cases in
figure~\ref{fig:pipf}. Incident rotons with energy $\varepsilon$ and $p>p_0$
(ordinary rotons) exhibit three different types of trajectory,
depending on the value of the angle of incidence.
Extraordinary rotons (with $p<p_0$) can only become ordinary. We have
$dT/dp = (p-p_0)/\mu$

\[p_x(t) = p_x(0)\]
\[p_y(t) = p_y(0) + ft\]
\[\dot{x}(t) = (p_x(0)/p(t))(p(t) - p_0)/\mu\]
\[\dot{y}(t) = (p_y(t)/p(t))(p(t) - p_0)/\mu\]

\begin{figure}[ht]
  \begin{center}
    \includegraphics{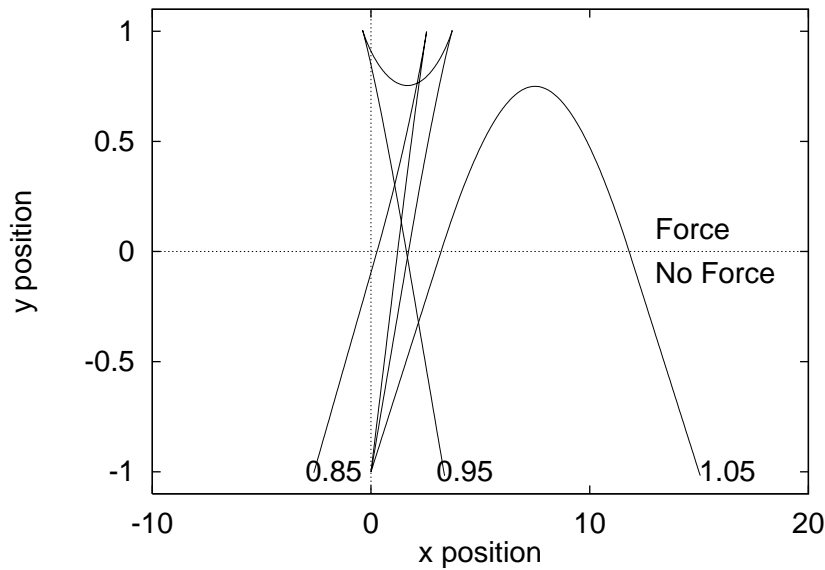}
  \end{center}
\caption{Coordinate space
  trajectories corresponding to momenta in the three regions shown in the
  previous figure. Notice the different scales in the two
  directions.}\label{fig:cftrajs}
\end{figure}

We show in figure~\ref{fig:cftrajs} coordinate space trajectories derived
from these results. There is a constant repulsive force in the region
$y > 0,$ no force for $y < 0$. A cusp appears in the trajectory when
$p$ becomes equal to $p_0,$ at which point the velocity is zero, but
the momentum vector is not. The unit of distance has been taken as the
initial ``kinetic'' energy divided by the magnitude of the force, so
that the roton stops after penetrating unit distance into the region.
Three $R^+$ rotons with momentum $1.1p_0$ have been started at the
point (0, -1) with tangential ($x$) momenta 1.05, 0.95, and 0.85 times
$p_0$. The trajectory of an $R^-$ roton is similar to the last of
these, but traversed in the opposite direction. The trajectories of
rotons with different values of the incident momentum would be similar
to these, differing only in a scale factor in the $x$ direction.

Such cusped trajectories were apparently first noted by Goodman
\cite{Goodman:1971} in analyzing roton trajectories near a vortex
line, a problem which we will discuss in section~\ref{sec:vortex}.
Note, however, that these ``snap back'' trajectories (an apt term used
in ref.~\cite{Goodman:1971}) have momentum transfer $\delta p$ of
order

\begin{eqnarray}
\delta p & \approx & p - \tilde{p} \nonumber \\
 & = & 2(p-p_0), \nonumber
\end{eqnarray}
a quantity which is small compared with $2p_0$. This reduces
significantly the pressure exerted on a wall by a roton gas.
This point has been emphasized by Maris and Cline~\cite{Maris:1981}.

\subsection{Pressure}

The results of the previous section imply that the pressure exerted by
a roton gas must be much below what one would na\"{\i}vely estimate
from the flux and a momentum transfer of order $p_0$. As emphasized by
Maris and Cline~\cite{Maris:1981}. the momentum transfer \emph{must} be
low to agree with the exact result, given in equation~\ref{p_rho}.

\[
 p_\rho =-(\partial{F_\rho }/\partial{V})_T=n_\rho k_BT.
\]

In the semiclassical case, considered here, the $R^-$ rotons always
change mode, never reflecting specularly; the $R^+$ rotons do the
same, except for nearly glancing angles. Summing the contributions to
the pressure from these various processes does yield the correct
result.

In quantum mechanics ``Whatever is not forbidden is allowed.'' In
particular, specular reflection is always consistent with the
conservation laws, even when it does not occur classically. When we
consider the branching to the other possible final state, in quantum
mechanics, it can not change the result. The pressure is independent
of the details of the processes at the walls. We can see explicitly
that there will be cancelation between specular reflection of $R^+$
rotons, which has large positive momentum transfer, and that of $R^-$
rotons, which produces a similar \emph{negative} contribution to the
pressure.

\section{Rotons at a free surface}
\label{sec:surface}

In an early paper which considered ballistic rotons,
P.~W.~Anderson~\cite{Anderson:1969} pointed out that, since the
minimum roton energy (8.6\,K) is greater than the binding energy of an
atom in the liquid ($E_B=7.2$\,K), a roton incident on the surface might
eject an atom into the vapor.

If the process is elastic, so that all the energy is transferred
between the elementary excitation and the atom, and the surface is
smooth, so that the component of momentum parallel to the surface is
conserved, then we can define an ``index of refraction'' in the usual
way. (Note, however, that an $R^-$ roton and an atom would have
\emph{oppositely} directed tangential velocity.)
\[ \tilde{n} = p_A/p = \sin\theta/\sin\phi, \]
where $p$ and $\theta$ are the momentum and angle of incidence of the
elementary excitation and $p_A$ and $\phi$ the corresponding
quantities for the helium atom. In this section, in order to connect
with experiments described in section~\ref{sec:expers}, we consider the
entire excitation spectrum---not just the roton region.

\begin{figure}[htbp]
  \begin{center}
    \includegraphics{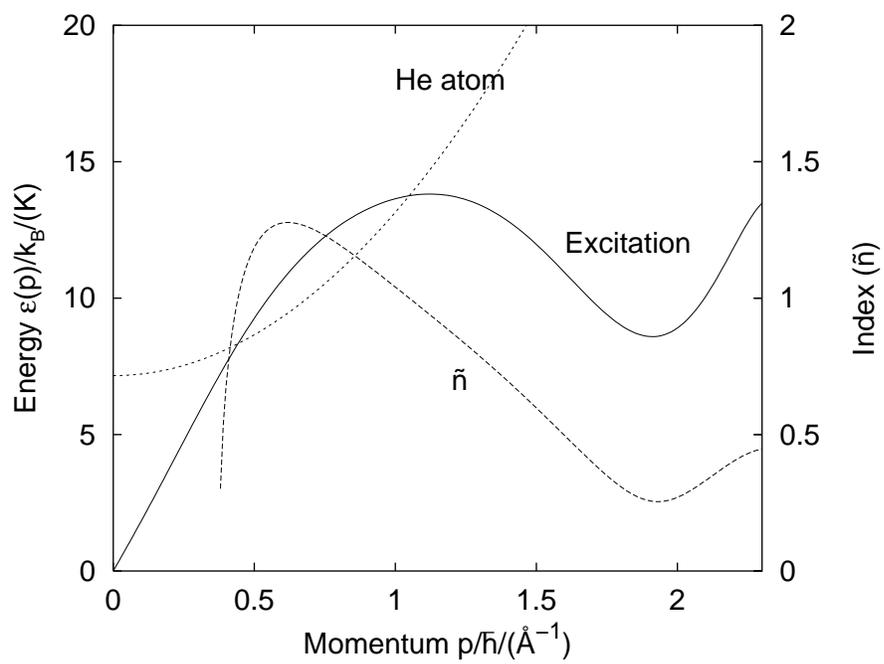}
    \caption{Energy of an
      elementary excitation, and an atom outside the liquid, and the
      ``Refractive index'' plotted against momentum/\,$\hbar$.}
    \label{fig:en-index}
  \end{center}
\end{figure}

The index $\tilde{n}$ is a single-valued function of $p,$ but a
multiple-valued function of $p_A,$ since an incident atom might produce
as many as three different excitations with a given energy. Written in
terms of $p$ (or $q = p/\hbar$), we have
\[\tilde{n} = \sqrt{(2m_4/p^2)(\varepsilon(p)-E_B)}\]
Figure~\ref{fig:en-index} plots the refractive index versus momentum, and
also shows the energy of the elementary excitation and a free atom of
equal momentum. The index equals 1 where the two energies cross.

\section{Roton-vortex interaction}
\label{sec:vortex}
\subsection{Quantized vortices}

The spatially-varying superfluid velocity field near a straight vortex
line is:

\begin{equation}
\label{vofr}
\vec{v}_s(\vec{r}) = \vec{\kappa} \times \vec{r} /2\pi r^2 = ( \kappa / 2 \pi r
) \hat{\varphi}.
\end{equation}
Here $\vec{r}$ is the two-dimensional (polar) radius vector, $\varphi$ the
polar angle, and $\vec{\kappa}$ is along the vortex line ($z$ direction).
The definition of the circulation $\kappa$ is
\[
\kappa \equiv \oint {\vec{v_s} \cdot \vec{dl}}
\]
and, in the case of superfluid liquid ${}^4\textrm{He}$, is quantized
in units of $h/m_4$. These equations are valid outside a core region,
which is of atomic dimensions.

\subsection{Quasiparticle-vortex interaction}

Goodman~\cite{Goodman:1971} and Samuels and
Donnelly~\cite{Samuels:1990} calculated the ``mutual friction''
between normal and superfluid using the roton-vortex interaction,
which causes momentum transfer between the rotons (a constituent of
the normal fluid) and the superfluid vortices.

If a quasiparticle is not too close to the vortex core, the velocity field
will not vary very rapidly, and we can combine
equations~\ref{pdotv}~and~\ref{vofr} to write:
\[\delta \varepsilon (\vec{r}) = \vec{p} \cdot \vec{\kappa} \times
\vec{r} /2\pi r^2, \]
which can be rewritten in the form
\begin{eqnarray}
\delta \varepsilon (\vec{r}) &=& \vec{\kappa} \cdot \vec{r} \times
\vec{p} /2\pi r^2 \nonumber \\
&=& \kappa L_z /2\pi r^2, \nonumber
\end{eqnarray}
if we take the $z$ axis along the vortex line. In this form, it is
apparent that $L_z,$ the $z$ component of angular momentum will be a
constant of the motion, permitting analysis of the trajectory to be
reduced to quadrature.~\cite{Sanders:1991}

\begin{figure}[htbp]
  \begin{center}
    \includegraphics{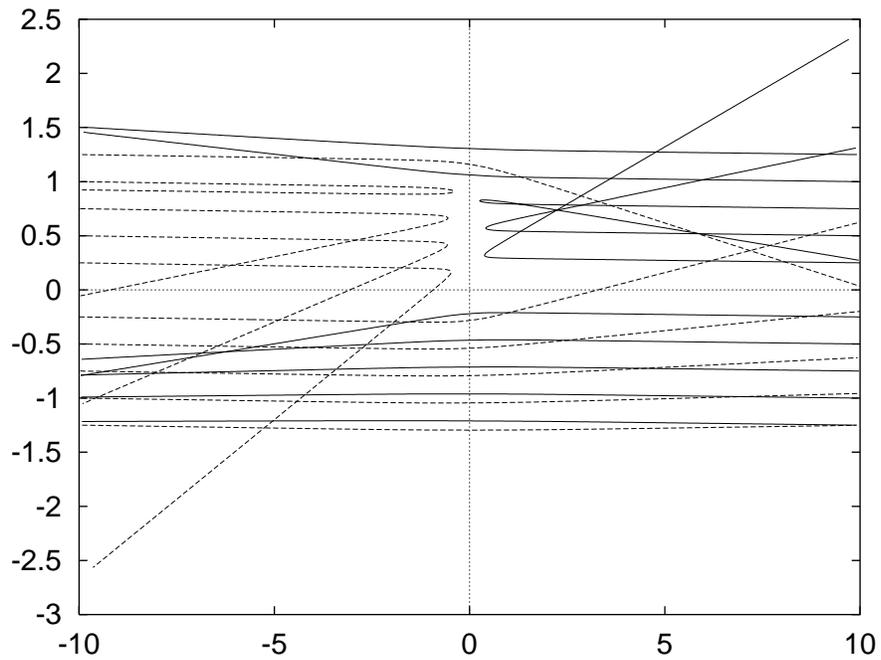}
    \caption{Calculated roton
      trajectories near a quantized vortex line, located at the
      origin. The curves are calculated for an incident $R^+$ roton
      with momentum $1.1p_0$ incident from the right (solid lines),
      and an $R^-$ roton with momentum $0.9p_0$ incident from the left
      (dashed lines).}
    \label{fig:rvtraj}
  \end{center}
\end{figure}

We show trajectories of $R^+$ and $R^-$ rotons, calculated by
numerically integrating the equations of motion, in
figure~\ref{fig:rvtraj}. An $R^+$ roton with incident $x$ momentum
$(p_i)_x=-1.1p_0$ is started from the right, an $R^-$ roton with $x$
momentum $(p_i)_x=0.9p_0$ from the left for various values of the
impact parameter. Distances are measured in units of a ``critical''
distance
\[ R^c = (\kappa/2\pi)(\mu/p_0)\frac{2p_i/p_0}{[(p_i/p_0) - 1]^2} =
(1/q_0)(\mu/m_4)\frac{2p_i/p_0}{[(p_i/p_0) - 1]^2},\]
at which the roton-vortex interaction energy is equal to the initial
roton kinetic energy, about 16\AA\ for the rotons illustrated.

Both the $R^+$ roton traveling to the left and the $R^-$ roton
traveling right have their momenta directed to the right. Thus, both
experience a repulsive interaction \emph{above} the (counter-clockwise
circulating) vortex, where $\vec{p}\cdot\vec{v_s}$ is positive. The
mode-changing trajectories all have impact parameters on the same
side of the vortex.

\section{The experiments}
\label{sec:expers}

Several experimenters performed pioneering experiments involving rotons
at temperatures low enough so that the roton mean free path is
long.

At Ohio State University, evidence for roton production was
sought in a study of helium atom reflection from the free
surface.~\cite{Edwards:1975}

In Paris, the inverse experiment was performed.~\cite{Balibar:1977}
Ballistic rotons, produced at a pulsed heater `evaporated' atoms from
the free surface.

These experiments have been extended and elaborated in a series of
experiments at University of Exeter.  In the Exeter experiments,
rotons (and phonons) are produced at a pulsed heater and detected by a
superconducting bolometer. The experiments are performed at
temperatures sufficiently low that the rotons may considered to be
collision-free (after they leave the vicinity of the source), and to
follow ``ballistic'' trajectories. We cite only a few of the articles
here, and refer the reader to these articles for additional
references. A rather comprehensive early article is.~\cite{Brown:1980}

The experiments include detection and measurement of roton-roton
scattering, produced by crossing two beams.~\cite{Forbes:1990b}

A puzzling aspect of these experiments is the observation that only $R^+$
rotons are produced at the heater, although there was evidence that
$R^-$ rotons could be produced at surfaces.

\subsection{Direct detection of $R^-$ rotons}
\label{sec:Rminus detection}

A compelling demonstration of the odd kinematic properties of the
$R^-$ roton would be evidence of oppositely-directed momentum and
velocity.

Tucker and Wyatt have recently realized a source of $R^-$
rotons,~\cite{Tucker:1998c} using two facing heaters, and utilizing
scattering of $R^+$ rotons to produce $R^-$ ones.  They have directed
these rotons at the free surface of the liquid, producing evaporation
of single atoms. As noted in section~\ref{sec:surface} the $R^-$ roton
and the atom will have \emph{oppositely} directed tangential velocity.

\begin{figure}[htbp]
  \begin{center}
    \includegraphics{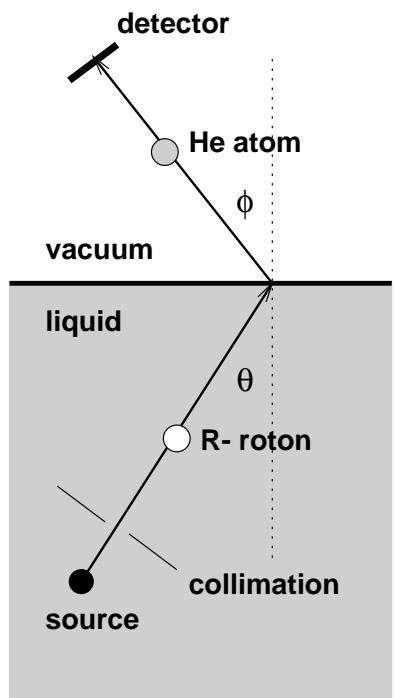}
    \caption{Evaporation of a He atom by an $R^-$ roton.}
    \label{fig:r-atom}
  \end{center}
\end{figure}

This type of confirmation of the anomalous connection between velocity
and momentum of the $R^-$ rotons~\cite{Tucker:1999}~\cite{Tucker:1998}
is accomplished  in the experiment pictured in
Fig.~\ref{fig:r-atom}.

The evaporated He atom is detected in the ``backward'' direction,
although the momentum component parallel to the surface is conserved in
the transition from roton to evaporated atom.

\appendix

\section{Appendix}

\subsection{Excitations in a moving fluid}
\label{sec:doppler}

We present a rather pedestrian but (we hope) straightforward derivation of
this familiar result. If we first think classically of viewing a system from
a reference system in which it is moving at velocity $\vec{v},$ each
particle will have $m\vec{v}$ added to its momentum, which will change the
momentum and kinetic energy, but for a system which is
translation-invariant (The interactions depend only on \emph{relative}
positions.), the potential energy will not change.

The same statements can be carried over to quantum mechanics. We can relate
the wave function of our system at rest to the wave function when it has
velocity $\vec{v}$ by

\[
 \Psi ^{(\vec{v})} = \exp\left[(mi/ \hbar ) \sum_{i=1}^N \vec{v} \cdot
 \vec{r_i}\right] \Psi , 
\]
where the index $i$ ranges over the $N$ atoms of the system. Since the
velocity enters only the phase factor, the expectation value of the
potential energy will not change.

The momentum and energy eigenvalues of the system in the two reference
frames are:
\begin{eqnarray}
\vec{P} ^{(\vec{v})} &=& \vec{P} + Nm \vec{v} \nonumber \\
E ^{(\vec{v})} &=& E + \vec{P} \cdot \vec{v} + Nmv^2 / 2.\nonumber 
\end{eqnarray}
Corresponding expressions for the ground state are:
\begin{eqnarray}
\vec{P}_0 ^{(\vec{v})} &=& Nm \vec{v}\nonumber \\
E_0 ^{(\vec{v})} &=& E_0 + Nmv^2 / 2. \nonumber
\end{eqnarray}
For the state in which a single quasiparticle $(\vec{p},\varepsilon)$ is
present: 
\begin{eqnarray}
\vec{P}_p ^{(\vec{v})} &=& \vec{p} + Nm \vec{v} \nonumber \\
E_{\vec{p}} ^{(\vec{v})} &=& E_0 + \varepsilon({\vec{p}}) + \vec{p} \cdot
\vec{v} + Nmv^2 / 2. \nonumber
\end{eqnarray}
Thus the change in momentum and energy associated with the single
quasiparticle is 
\begin{eqnarray}
\vec{p}^{(\vec{v})} &=& \vec{P}^{(\vec{v})}_{\vec{p}} -
\vec{P}^{(\vec{v})}_0 \nonumber \\ 
 &=& \vec{p} \nonumber \\
\varepsilon ^{(\vec{v})} (\vec{p}) &=& E_{\vec{p}} ^{(\vec{v})} - E_0
^{(\vec{v})} \nonumber \\ 
 &=& \varepsilon (\vec{p}) + \vec{p} \cdot \vec{v}, \nonumber
\end{eqnarray}
which is our equation~\ref{doppler} of section~\ref{sec:vel}

\newpage
\noindent
\textbf{\Large Figure captions}
\vspace{0.125in}
\begin{description}
\item [Figure~\ref{fig:landaucurve}]Dispersion curve for elementary
  excitations in liquid helium, calculated from the polynomial fit in
  reference~{\protect \cite{Brooks:1977}}. The roton region is near
  the minimum.
\item [Figure~\ref{fig:isoenergy}]Surfaces of constant energy
  in momentum space for $\varepsilon=\Delta$ (dotted), and a higher
  energy (solid).
\item [Figure~\ref{fig:pipf}]Momentum space diagram
  showing initial and final momenta corresponding to reflection of
  rotons by a force in the $-y$ direction.
\item [Figure~\ref{fig:cftrajs}]Coordinate space
  trajectories corresponding to momenta in the three regions shown in the
  previous figure. Notice the different scales in the two
  directions.
\item [Figure~\ref{fig:en-index}]Energy of an elementary excitation,
  and an atom outside the liquid, and the ``Refractive index'' plotted
  against momentum/\,$\hbar$.
\item [Figure~\ref{fig:rvtraj}]Calculated roton
      trajectories near a quantized vortex line, located at the
      origin. The curves are calculated for an incident $R^+$ roton
      with momentum $1.1p_0$ incident from the right (solid lines),
      and an $R^-$ roton with momentum $0.9p_0$ incident from the left
      (dashed lines).
\item [Figure~\ref{fig:r-atom}]Evaporation of a He atom by an $R^-$ roton.
\end{description}
\newpage
\bibliography{lhe}
\bibliographystyle{unsrt}
\end{document}